\documentclass{elsart}
\usepackage{epsfig}
\begin{document}
\begin{flushright}
KUPT 02-03 \\
June 2002
\end{flushright}
\begin{frontmatter}
\title{Consequences of symmetries \\
in renormalizing collinear effective theory} 
\author{Junegone Chay\thanksref{email}},
\thanks[email]{E-mail address: chay@korea.ac.kr}
\author{Chul Kim}
\address{Department of Physics, Korea University, Seoul 136-701,
Korea} 
\begin{abstract}
We consider effects of symmetries on renormalization properties of
the collinear effective theory. We investigate which types of
operators are possible in the effective theory satisfying gauge
invariance, reparameterization invariance and residual energy
invariance. Each symmetry puts a constraint on the possible structure
of the theory, and there can appear only specific combinations of
operators in the effective Lagrangian satisfying all the symmetry
requirements. And the final effective Lagrangian is not renormalized
to all orders in $\alpha_s$ as long as no other nonlocal operators are
induced at higher order. We explicitly prove this at one loop by
renormalizing one-gluon vertices and discuss their features.  
\end{abstract}
\end{frontmatter}

Strong interaction processes which involve energetic, massless
particles can be described by the collinear effective theory
\cite{bauer1,bauer2,bauer3,bauer4}. 
It has been applied to sum Sudakov
logarithms \cite{bauer1}, to prove factorization \cite{bauer5,bauer6}
and study power corrections \cite{chay1}. Symmetry properties such as
reparameterization invariance, residual energy invariance, and gauge
invariance have been investigated \cite{manohar,chay2}.

The collinear effective theory  offers a systematic
way to organize physical quantities in powers of a small parameter
$\lambda \sim p_{\perp}/E$, in which a massless energetic quark moves
with energy $E$, and transverse momentum $p_{\perp}$. The momentum
$P^{\mu}$ of an energetic particle can be decomposed as
\begin{equation}
  \label{eq:momentum}
P^{\mu} = \frac{\overline{n} \cdot p}{2} n^{\mu} +p_{\perp}^{\mu}
+k^{\mu},
\end{equation}
where $p^{\mu} = \half (\overline{n}\cdot p)n^{\mu} + p_{\perp}^{\mu}$
is the label momentum of order $\lambda^0$ and $\lambda$, respectively.
The momentum $k^{\mu}$ is the residual momentum of order $\lambda^2$,
which represents small fluctuation due to the strong interaction.

In the collinear effective theory, we classify fields into three
classes according to their momenta as collinear, soft and
ultrasoft (usoft) fields. Their typical momenta scale as
$E(\lambda^2,1,\lambda)$, $E(\lambda,\lambda,\lambda)$, and
$E(\lambda^2, \lambda^2, \lambda^2)$, respectively. In processes
involving collinear quarks, the relevant fields are collinear quarks
$\xi_n$, collinear gluons $A_n^{\mu}$ and usoft gluons $A_u^{\mu}$. 
The effective Lagrangian can be derived from the full QCD in terms of
the collinear quark spinor $\xi_n$ which satisfies
\begin{equation}
  \label{eq:xin}
\frac{\FMslash{n} \overline{\FMslash{n}}}{4} \xi_n =\xi_n, \
\FMslash{n} \xi_n=0,
\end{equation}
where $n^2 =0$, $\overline{n}^2=0$ and $n\cdot \overline{n}=2$. The
effective Lagrangian is written as
\begin{equation}
  \label{eq:eff}
\mathcal{L} =  \overline{\xi}_n \Bigl[ n\cdot i\mathcal{D} 
+i\FMslash{\mathcal{D}}_{\perp}
\frac{1}{\overline{n} \cdot i\mathcal{D}}
i\FMslash{\mathcal{D}}_{\perp}
\Bigr] \frac{\FMslash{\overline{n}}}{2} \xi_n,
\end{equation}
where $\mathcal{D}^{\mu} = D_c^{\mu} + D_u^{\mu}$ is a covariant
derivative under collinear and usoft gauge transformations. The
covariant derivatives $D_c$ and $D_u$ are defined as
\begin{equation}
  \label{eq:cov}
iD_c^{\mu} = \mathcal{P}^{\mu} -gA_n^{\mu}, \ \
iD_u^{\mu} = i\partial^{\mu} -gA_u^{\mu}.   
\end{equation}
Here $\mathcal{P}^{\mu}$ is the operator which extracts label momenta
from collinear fields. For example, if we apply $\mathcal{P}^{\mu}$ to
a collinear spinor $\xi_n$ with label momentum $p^{\mu}$, we get
\begin{equation}
\mathcal{P}^{\mu} \xi_n = \Bigl(\frac{\overline{n}\cdot p}{2} n^{\mu}
+p_{\perp}^{\mu} \Bigr) \xi_n.
\end{equation}
Note that the derivative operator acting on a collinear field produces
terms of order $\lambda^2$ since the label momenta are extracted. 

At leading order in $\lambda$, Manohar et al. \cite{manohar} have
observed that, if we require only gauge invariance and
reparameterization invariance, there can be operators in the
effective Lagrangian, which are given by 
\begin{equation}
  \label{eq:op0}
O_2^{(0)} = \overline{\xi}_n i\FMslash{D}_{\perp c}
\frac{1}{\overline{n}\cdot iD_c} i\FMslash{D}_{\perp c}
\frac{\overline{\FMslash{n}}}{2} \xi_n, \ \  
O_3^{(0)} = \overline{\xi}_n iD_{\perp c\mu}
\frac{1}{\overline{n}\cdot iD_c} iD_{\perp c}^{\mu}
\frac{\overline{\FMslash{n}}}{2} \xi_n.
\end{equation}
Even though only $O_2^{(0)}$ appears at tree level, $O_3^{(0)}$ can be
induced through radiative corrections. They have shown that 
the operator $O_3^{(0)}$ at leading order in $\lambda$ is
not allowed due to the type-II reparameterization invariance,
though the type-I reparameterization invariance does not exclude the
operator. In this Letter, we extend the analysis to all orders in
$\lambda$ by including all the possible operators in the collinear
effective theory. We also present explicit calculations for the
renormalization of one-gluon vertices at one loop to show that the
effective Lagrangian in a specific combination is not renormalized at
this order.  

We can obtain the operators similar to $O_2^{(0)}$ and $O_3^{(0)}$ to
all orders. If we only require the  collinear and usoft
gauge invariance \cite{chay2}, the most general set of the operators
in the effective Lagrangian is given by 
\begin{eqnarray}
  \label{eq:decom}
O_1 &=& \overline{\xi}_n n\cdot
i\mathcal{D}\frac{\overline{\FMslash{n}}}{2} \xi_n,  \ 
O_2 = \overline{\xi}_n i\FMSlash{\mathcal{D}}_{\perp}
\frac{1}{\overline{n} \cdot i\mathcal{D}}
i\FMSlash{\mathcal{D}}_{\perp} \frac{\overline{\FMslash{n}}}{2} \xi_n,
\nonumber \\
O_3 &=& \overline{\xi}_n i\mathcal{D}_{\perp \mu}
\frac{1}{\overline{n}  \cdot i\mathcal{D}} i\mathcal{D}_{\perp}^{\mu}
\frac{\overline{\FMslash{n}}}{2} \xi_n, \
O_4 =  \overline{\xi}_n (-i\sigma^{\mu\nu}) 
i\mathcal{D}_{\perp \mu} \frac{1}{\overline{n}  \cdot i\mathcal{D}}
i\mathcal{D}_{\perp\nu} \frac{\overline{\FMslash{n}}}{2} \xi_n,      
\end{eqnarray}
where $\sigma^{\mu\nu} = i[\gamma^{\mu},\gamma^{\nu} ]/2$. Note that the
operators $O_2$, $O_3$ and $O_4$ are not independent and the relation
is given by $O_2 = O_3 +O_4$ due to the identity 
$\gamma^{\mu} \gamma^{\nu} = g^{\mu \nu} -i\sigma^{\mu\nu}$.

At tree level there only appears $O_1+O_2$, but the operator $O_2$ can
be decomposed into $O_3+O_4$. And the question is whether the operators
$O_3$ and $O_4$ receive different renormalization effects. To compare
this situation with the heavy quark effective theory (HQET), the
effective Lagrangian to order $1/m_Q$ in HQET is given by
\begin{eqnarray}
  \label{eq:hqet}
\mathcal{L}_{\mathrm{HQET}} &=& \overline{h}_v iv\cdot D h_v
+\overline{h}_v \frac{(i\FMSlash{D})^2}{2m_Q} h_v \nonumber \\ 
&=&  \overline{h}_v iv\cdot D h_v +\overline{h}_v \frac{(iD)^2}{2m_Q}
h_v - C_{\mathrm{mag}} (\mu) 
\frac{g}{4m_Q} \overline{h}_v \sigma_{\mu\nu} G^{\mu\nu} h_v,
\end{eqnarray}
where $iD^{\mu} = i\partial^{\mu} -gA^{\mu }$. At one loop, when
we match the full theory with the HQET, the coefficient of the
chromomagnetic operator $C_{\mathrm{mag}} (\mu)$ is given by
\cite{eichten,falk} 
\begin{equation}
C_{\mathrm{mag}} (\mu) = 1-\frac{3\alpha_s}{2\pi} \Bigl( \ln
\frac{m_Q}{\mu} -\frac{13}{9} \Bigr).
\end{equation}
The last two operators in Eq.~(\ref{eq:hqet}) are legitimate operators
which are gauge invariant. However, when we include radiative
corrections, the kinetic energy operator is not renormalized due to the
reparameterization invariance, but the chromomagnetic operator
receives nontrivial renormalization effects.  

If we follow the same reasoning in the HQET, we could conclude that
the kinetic energy operator $O_1+O_3$ is not
renormalized to all orders in $\alpha_s$. However, we can extend the
reasoning further by including the residual energy invariance. The
residual energy invariance along with the reparameterization
invariance guarantees that only the combination $O_1+O_3 +
O_4=O_1+O_2$ can appear in the Lagrangian, and is not renormalized to
all orders.  

In order to see how the argument goes, let us consider the
transformation properties of each operator 
under the reparameterization transformation and the residual energy
transformation. These transformations are classified in
Ref.~\cite{manohar} as the type-I and the type-II transformations
which are given by
\begin{equation}
  \label{eq:genrep}
(\mathrm{I}) \left\{ \begin{array}{l}
n^{\mu}  \rightarrow n^{\mu} + \Delta^{\mu}_{\perp}, \\
\overline{n}^{\mu} \rightarrow \overline{n}^{\mu},
\end{array}
\right. \  (\mathrm{II}) \left\{ \begin{array}{l}
n^{\mu}  \rightarrow n^{\mu}, \\
\overline{n}^{\mu} \rightarrow
\overline{n}^{\mu}+\epsilon^{\mu}_{\perp},
\end{array}
\right.
\end{equation}
where the infinitesimal parameters $\Delta_{\perp}^{\mu}$ and
$\epsilon_{\perp}^{\mu}$ satisfy $n\cdot \Delta_{\perp} = \overline{n}
\cdot \Delta_{\perp} = n\cdot \epsilon_{\perp} = \overline{n} \cdot
\epsilon_{\perp}=0$. 
Under the type-I transformation, the transformation of each quantity 
in the Lagrangian is given by
\begin{eqnarray}
\label{eq:type1}
n\cdot \mathcal{D} &\rightarrow& n\cdot \mathcal{D} +\Delta_{\perp}
\cdot \mathcal{D}_{\perp}, \
\mathcal{D}_{\perp}^{\mu}  \rightarrow  \mathcal{D}_{\perp}^{\mu}
-\frac{\Delta_{\perp}^{\mu}}{2} \overline{n} \cdot \mathcal{D}
-\frac{\overline{n}^{\mu}}{2} \Delta_{\perp}\cdot \mathcal{D}_{\perp},
\nonumber \\
\overline{n}\cdot \mathcal{D} &\rightarrow&
\overline{n}\cdot \mathcal{D}, \ \xi_n \rightarrow (1+\frac{1}{4}
\FMSlash{\Delta}_{\perp} \overline{\FMslash{n}}) \xi_n.
\end{eqnarray}
Using Eq.~(\ref{eq:type1}), the infinitesimal change of the operators is
given by
\begin{equation}
\delta_{\mathrm{(I)}} O_1 = \overline{\xi}_n (i \Delta_{\perp}\cdot
\mathcal{D}_{\perp}) \frac{\overline{\FMslash{n}}}{2} \xi_n =
-\delta_{\mathrm{(I)}} O_2 = -\delta_{\mathrm{(I)}} O_3, \
\delta_{\mathrm{(I)}} O_4 =0. 
\end{equation}
Therefore the combinations $O_1 +O_2$, $O_1+O_3$ and $O_4$ are
invariant under the reparameterization (type-I) transformation. 

Under the type-II transformation, each vector and the spinor $\xi_n$
transform as 
\begin{eqnarray}
n\cdot \mathcal{D} &\rightarrow& n\cdot \mathcal{D}, \
\mathcal{D}_{\perp}^{\mu}  \rightarrow  \mathcal{D}_{\perp}^{\mu}
-\frac{\epsilon_{\perp}^{\mu}}{2} n \cdot \mathcal{D}
-\frac{n^{\mu}}{2} \epsilon_{\perp}\cdot \mathcal{D}_{\perp},
\nonumber \\
\overline{n}\cdot \mathcal{D} &\rightarrow&
\overline{n}\cdot \mathcal{D}+\epsilon_{\perp} \cdot
\mathcal{D}_{\perp}, \ \xi_n \rightarrow (1+\frac{1}{2}
\FMslash{\epsilon}_{\perp}\frac{1}{\overline{n}\cdot \mathcal{D}}
\FMslash{\mathcal{D}}_{\perp}) \xi_n.
\end{eqnarray}
From this, we obtain the change of $O_1$, which is given by 
\begin{equation}
  \label{eq:lamlag}
\delta_{(\mathrm{II})} O_1= \overline{\xi}_n \Bigg(
i\FMSlash{\mathcal{D}}_{\perp} \frac{1}{\overline{n} \cdot
    i\mathcal{D}} \frac{\FMslash{\epsilon}_{\perp}}{2} n\cdot
  i\mathcal{D} + n\cdot i\mathcal{D}
  \frac{\FMslash{\epsilon}_{\perp}}{2} \frac{1}{\overline{n} \cdot
    i\mathcal{D}} i\FMSlash{\mathcal{D}}_{\perp} \Bigg)
    \frac{\overline{\FMslash{n}}}{2} \xi_n.  
\end{equation}
The variation of $O_2$ exactly cancels the change in $O_1$ and is
given by 
\begin{equation}
  \label{eq:varo1}
\delta_{(\mathrm{II})} O_2 = -\delta_{(\mathrm{II})} O_1.
\end{equation}

The change of $O_3$ and $O_4$ is given by
\begin{eqnarray}
  \label{eq:varo2}
\delta_{(\mathrm{II})} O_3 &=& \overline{\xi}_n
\Bigg[i\FMSlash{\mathcal{D}}_{\perp} \frac{1}{\overline{n} \cdot
  i\mathcal{D}} \frac{\FMslash{\epsilon}_{\perp}}{2}
i\mathcal{D}_{\perp\mu}\frac{1}{\overline{n} \cdot
  i\mathcal{D}}i\mathcal{D}_{\perp}^{\mu}  -n\cdot
i\mathcal{D}\frac{1}{\overline{n} \cdot   i\mathcal{D}}
\frac{\epsilon_{\perp} \cdot i\mathcal{D}_{\perp}}{2}   \nonumber \\
&&- i\mathcal{D}_{\perp\mu}\frac{1}{\overline{n} \cdot
  i\mathcal{D}} \epsilon_{\perp}\cdot
i\mathcal{D}_{\perp} \frac{1}{\overline{n} \cdot i\mathcal{D}}
i\mathcal{D}_{\perp}^{\mu}  -  \frac{\epsilon_{\perp} \cdot
  i\mathcal{D}_{\perp}}{2} \frac{1}{\overline{n} \cdot
  i\mathcal{D}} n\cdot i\mathcal{D} \nonumber \\
&&+ i\mathcal{D}_{\perp\mu} \frac{1}{\overline{n} \cdot
  i\mathcal{D}} i\mathcal{D}_{\perp}^{\mu}
\frac{\FMslash{\epsilon}_{\perp}}{2}  \frac{1}{\overline{n} \cdot
i\mathcal{D}}   i\FMSlash{\mathcal{D}}_{\perp} \Bigg]
 \frac{\overline{\FMslash{n}}}{2} \xi_n, \nonumber \\
\delta_{(\mathrm{II})} O_4 &=& \overline{\xi}_n 
i\FMSlash{\mathcal{D}}_{\perp} \frac{1}{\overline{n} \cdot
  i\mathcal{D}} \frac{\FMslash{\epsilon}_{\perp}}{2} (-i\sigma^{\mu\nu})
i\mathcal{D}_{\perp\mu}\frac{1}{\overline{n} \cdot
  i\mathcal{D}}i\mathcal{D}_{\perp\nu}
\frac{\overline{\FMslash{n}}}{2} \xi_n \nonumber \\
&&+\overline{\xi}_n  (-i\sigma^{\mu\nu}) \Bigg[
-n\cdot
i\mathcal{D}\frac{1}{\overline{n} \cdot   i\mathcal{D}} 
\frac{\epsilon_{\perp\mu} i\mathcal{D}_{\perp\nu}}{2}  \nonumber \\
&&- i\mathcal{D}_{\perp\mu}\frac{1}{\overline{n} \cdot
  i\mathcal{D}} \epsilon_{\perp}\cdot
i\mathcal{D}_{\perp} \frac{1}{\overline{n} \cdot i\mathcal{D}}
i\mathcal{D}_{\perp\nu} 
-\frac{i\mathcal{D}_{\perp\mu}\epsilon_{\perp\nu}}{2} 
\frac{1}{\overline{n} \cdot 
  i\mathcal{D}} n\cdot i\mathcal{D} \nonumber \\
&&+ i\mathcal{D}_{\perp\mu} \frac{1}{\overline{n} \cdot
  i\mathcal{D}} i\mathcal{D}_{\perp\nu}
\frac{\FMslash{\epsilon}_{\perp}}{2}  \frac{1}{\overline{n} \cdot
i\mathcal{D}}   i\FMSlash{\mathcal{D}}_{\perp}  \Bigg]
 \frac{\overline{\FMslash{n}}}{2} \xi_n.
\end{eqnarray}
When we add the variations of $O_3$ and $O_4$, we can obtain the
relation 
\begin{equation}
  \label{eq:12com}
\delta_{(\mathrm{II})} (O_3+O_4) =  \delta_{(\mathrm{II})} O_2 =
-\delta_{(\mathrm{II})} O_1. 
\end{equation}
From this, we can conclude that the invariant combinations under the
type-II transformation are $O_1 + O_2$ and $O_1+O_3+O_4$, which are
the same. 

To summarize the result, the reparameterization invariance (type-I)
requires that $O_1$ should appear as a combination either of $O_1+O_2$
or $O_1+O_3$, and the operator $O_4$ itself is reparameterization
invariant. There is no prescription between $O_1$ and 
$O_4$. And the residual energy invariance (type-II) requires that
$O_1$ should appear as a combination either of $O_1+O_2$ or
$O_1+O_3+O_4$, which is the same as $O_1+O_2$. Therefore the residual
energy invariance puts a serious constraint on the structure of the
effective Lagrangian, and, as a consequence, the combination $O_1+O_2$
is not renormalized to all orders.  

We can explicitly prove that the combination $O_1+O_2$ is not
renormalized at one loop. To be concrete, let us 
consider the renormalization of the effective Lagrangian at leading
order in $\lambda$. In principle, we have to consider all the
operators, but here we consider the renormalization of one-gluon
vertices only to illustrate the point. The coefficients of the
operators in the collinear effective theory are determined by matching
amplitudes between the full theory and the effective theory. It is
convenient to use the background field method \cite{abbott} for an
external gluon with two external collinear fermions and we employ
the Feynman gauge in the calculation.

The Feynman diagrams for the one-gluon vertex to one loop in the full
theory are given in Fig.~1. We use the on-shell 
renormalization and regulate both infrared and ultraviolet divergences
using dimensional regularization with $D=4-2 \epsilon$. Because the
effective theory has the same low energy behavior as in the full
theory, the infrared divergences in the effective theory will exactly
cancel the infrared divergences in the full theory in
matching. Therefore one expects that the result will be independent of 
the choice of the infrared regulators. 

\begin{figure}[t]
\begin{center}
\epsfig{file=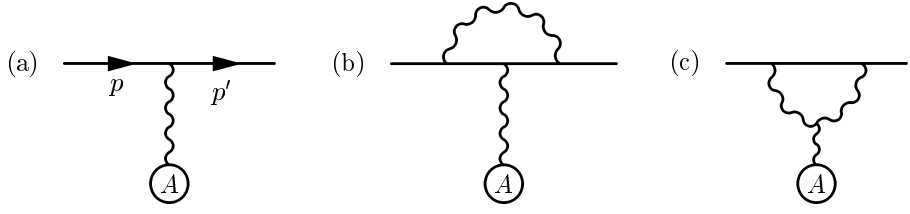}
\end{center}
\caption{Feynman diagrams of the full theory amplitudes for the
one-gluon vertex to one loop with the background field method.} 
\label{mag0}
\end{figure}

We can write the amplitude for the Feynman diagrams in Fig.~1 as
\begin{equation}
\label{fullcur}
-ig A_{\mu a} \overline{\psi} (p^{\prime}) \Gamma^{\mu} T^a \psi (p).
\end{equation}
The tree-level diagram which contributes to $\Gamma^{\mu}$ is simply
$\gamma^{\mu}$, and the Feynman diagrams at one-loop in Fig.~1 yield
\begin{equation}
-ig \frac{\alpha_s}{4\pi} C_F \gamma^{\mu} 
 \Bigl(\frac{1}{\epsilon_{\mathrm{UV}}}
 -\frac{1}{\epsilon_{\mathrm{IR}}} \Bigr),
\end{equation}
which is exactly cancelled by the wave function renormalization. This
is because the current is conserved.

We can find the interaction in the collinear
effective theory corresponding to  Eq.~(\ref{fullcur}) by noting that
the relation between the spinor $\psi$ in the full theory and the
spinor $\xi_n$ in the effective theory is given by
\begin{equation}
\psi (x) = \sum_p e^{-ip\cdot x} \Bigl(1+\frac{1}{\overline{n} \cdot
  i\mathcal{D}} i\FMSlash{\mathcal{D}}_{\perp}
  \frac{\overline{\FMslash{n}}}{2} \Bigr) \xi_n.
\end{equation} 
The interaction can be written as
\begin{equation}
-g\overline{\psi} \FMSlash{A} \psi \rightarrow -g\overline{\xi_n}
 \Bigl[ n \cdot A_n +i\FMSlash{\mathcal{D}}_{\perp}
 \frac{1}{i\overline{n} \cdot \mathcal{D}} \FMslash{A}_{n\perp}  +
 \FMslash{A}_{n\perp} \frac{1}{i\overline{n} \cdot \mathcal{D}}
 i\FMSlash{\mathcal{D}}_{\perp}\Bigr] \frac{\overline{\FMslash{n}}}{2}
 \xi_n,
\end{equation}
which gives the interaction term in the effective Lagrangian. 

\begin{figure}[t]
\begin{center}
\epsfig{file=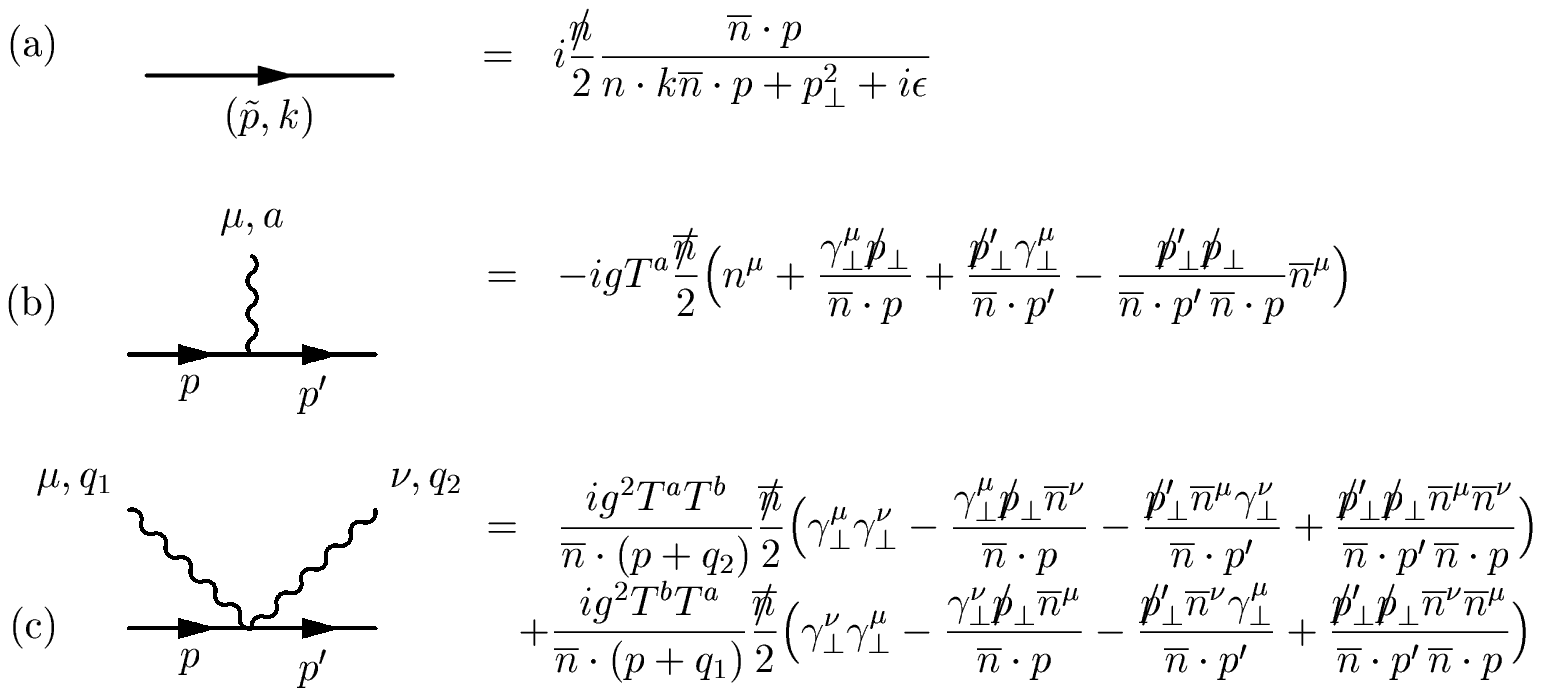,width=13.0cm}
\end{center}
\caption{Feynman rules for the propagator and the interaction
  vertices in the collinear effective theory. All the particles are
  collinear particles and the gluon momenta
  are incoming. }  
\label{rule}
\end{figure}

\begin{figure}[t]
\begin{center}
\epsfig{file=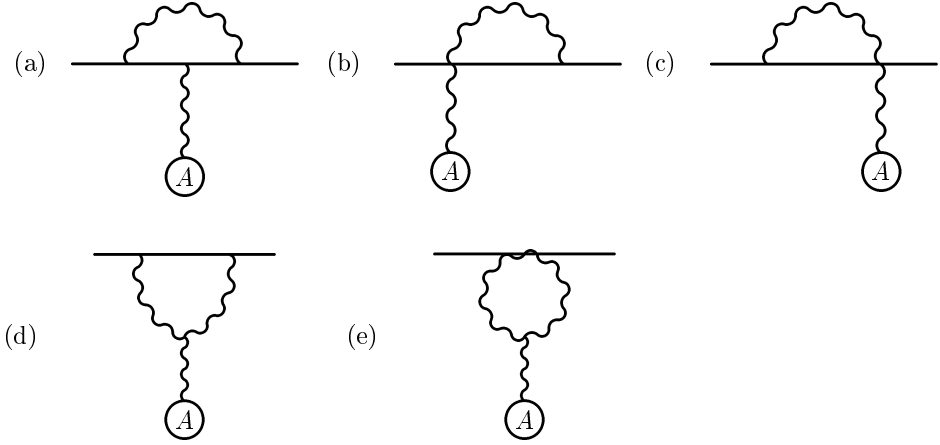}
\end{center}
\caption{Feynman diagrams for the one-gluon vertex in the effective
theory  with the background field method. The wavy lines represent
collinear gluons. Diagrams with usoft gluons, which do not contribute,
are not shown.} 
\label{mag1}
\end{figure}

The effective Lagrangian at leading order in $\lambda$ is given by
\begin{equation}
\mathcal{L}_0 = \overline{\xi}_n \Bigl[ n\cdot (iD-gA_n)
+(\FMSlash{\mathcal{P}}_{\perp} -g\FMSlash{A}_{\perp} )
\frac{1}{\overline{n} \cdot (\mathcal{P} -gA_n)}
(\FMSlash{\mathcal{P}}_{\perp} -g\FMSlash{A}_{\perp} )
\Bigr] \frac{\FMslash{\overline{n}}}{2} \xi_n.
\end{equation}
And the Feynman rules for the
interactions are given in Fig.~\ref{rule}. Here we omit vertices
with an usoft gluon, which do not contribute to the following
calculations. The relevant Feynman diagrams in the collinear effective
theory for the renormalization of a single-gluon
vertex are shown in Fig.~\ref{mag1}. Here we also use the background
field method for the external gluon field. There are also Feynman
diagrams with usoft gluons in Fig.~\ref{mag1}, but all of them
vanish. When we add the first three diagrams in Fig.~\ref{mag1},
the ultraviolet divergent part is given by
\begin{equation}
M_a + M_b + M_c = ig \frac{\alpha_s}{4\pi}\frac{T^a}{2N}
\frac{1}{\epsilon} \Bigl[ n_{\mu} +\frac{{\gamma}_{\perp\mu}
  \FMslash{p}_{\perp}}{\overline{n}\cdot p}
+\frac{\FMslash{p}^{\prime}_{\perp} \gamma_{\perp\mu}}{\overline{n}
  \cdot p^{\prime}} -\overline{n}_{\mu}
\frac{\FMslash{p}^{\prime}_{\perp}
  \FMslash{p}_{\perp}}{\overline{n}\cdot p^{\prime} \overline{n} \cdot
  p} \Bigr]\frac{\FMslash{\overline{n}}}{2}.
\label{cont1}
\end{equation}
If we regulate infrared and ultraviolet divergences using dimensional
regularization, the $1/\epsilon$ pole in Eq.~(\ref{cont1}) is replaced
by
\begin{equation}
  \label{eq:pole}
\frac{1}{\epsilon} \rightarrow \frac{1}{\epsilon_{\mathrm{UV}}}
-\frac{1}{\epsilon_{\mathrm{IR}}} 
\end{equation}
and the amplitude vanishes. In order to extract the ultraviolet
divergence, we can regulate the infrared divergence by
putting external particles slightly off the mass shell. Whatever
regularization method we use, the infrared divergences in the full
theory and the effective theory cancel in the matching.

The last two Feynman diagrams from a triple gluon vertex yield
\begin{equation}
M_d +M_e =  -ig \frac{\alpha_s}{4\pi} \frac{N}{2} T^a
\frac{1}{\epsilon} \Bigl[ n_{\mu} +\frac{{\gamma}_{\perp\mu}
  \FMslash{p}_{\perp}}{\overline{n}\cdot p}
+\frac{\FMslash{p}^{\prime}_{\perp} \gamma_{\perp\mu}}{\overline{n}
  \cdot p^{\prime}} -\overline{n}_{\mu}
\frac{\FMslash{p}^{\prime}_{\perp}
  \FMslash{p}_{\perp}}{\overline{n}\cdot p^{\prime} \overline{n} \cdot
  p} \Bigr]\frac{\FMslash{\overline{n}}}{2}.
\label{cont2}
\end{equation}
The contributions of all the Feynman diagrams in Fig.\ref{mag1} are
given by
\begin{equation}
M=  -ig \frac{\alpha_s}{4\pi} C_F T^a
\frac{1}{\epsilon} \Bigl[ n_{\mu} +\frac{{\gamma}_{\perp\mu}
  \FMslash{p}_{\perp}}{\overline{n}\cdot p}
+\frac{\FMslash{p}^{\prime}_{\perp} \gamma_{\perp\mu}}{\overline{n}
  \cdot p^{\prime}} -\overline{n}_{\mu}
\frac{\FMslash{p}^{\prime}_{\perp}
  \FMslash{p}_{\perp}}{\overline{n}\cdot p^{\prime} \overline{n} \cdot
  p} \Bigr]\frac{\FMslash{\overline{n}}}{2}.
\label{sum}
\end{equation}

The self-energy for a collinear quark is given by
\begin{equation}
i\Sigma (p) =i\frac{\alpha_s}{4\pi} C_F \frac{1}{\epsilon}
\frac{p^2}{\overline{n}\cdot p} \frac{\FMslash{\overline{n}}}{2},
\end{equation}
and the ultraviolet divergence is the same as that in the full
theory. When we add all these contributions, the ultraviolet divergent
part vanishes and we have shown that the one-gluon vertex operator in
$O_1+O_2$ or $O_1+O_3+O_4$ is not renormalized at one loop.

This result is expected since there are only massless particles both
in the full theory and in the effective theory. If we regulate both
the infrared divergence and the ultraviolet divergence using
dimensional regularization, any loop diagram either in the full
theory or in the effective theory vanishes
since there is no scale involved. Therefore matrix elements in each
theory are given by their tree-level expressions. Furthermore,
since the infrared divergence in the
full theory is the same as the infrared divergence in the effective
theory, the ultraviolet divergence in the effective theory should be
the same as the ultraviolet divergence in the full theory, in which
there is none in this case.

\begin{figure}[t]
\begin{center}
\epsfig{file=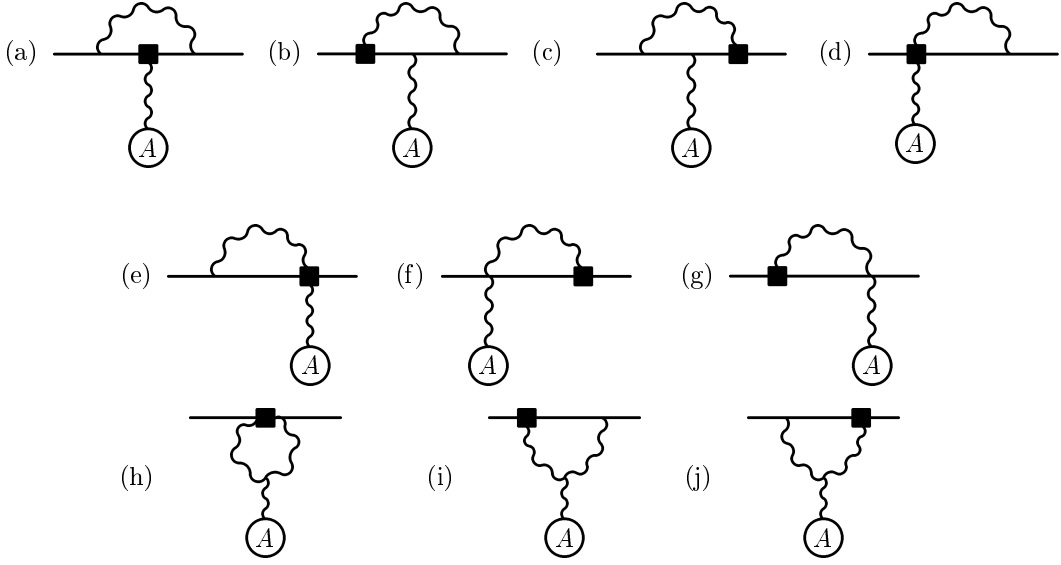}
\end{center}
\caption{Feynman diagrams at one loop in the collinear effective
  theory with the background field method in renormalizing the
  operator $O_4$. The square represents the vertex from the
  operator $O_4$. Wavy lines represent
  collinear gluons.}
\label{mag2}
\end{figure}

When we compare the renormalization behavior of the collinear
effective theory with that of the HQET, there is a distinct
difference.  
In HQET, when we calculate radiative corrections at one loop, the
kinetic energy operator and the chromomagnetic operator do not
mix. And the kinetic energy operator is not renormalized due to the
reparameterization invariance. However, in the collinear effective
theory, even though $O_1+O_3$ and $O_4$ have reparameterization
invariance, it is not guaranteed that radiative corrections of
$O_1+O_3$ and $O_4$ do not mix with each other. 

In order to see this,
let us consider the radiative corrections of $O_4$ at one loop. The
relevant Feynman diagrams are shown in Fig.~\ref{mag2}. It is not
illuminating to show all the results, but if we calculate the
radiative corrections of the operator $O_4$ in Fig.~\ref{mag2}, there
are terms proportional to $n \cdot A_n$, which are given by 
\begin{equation}
  \label{eq:nmu}
g \frac{\alpha_s}{4\pi} \frac{1}{\epsilon} \overline{\xi}_n \Bigl(
  \frac{1}{2N} 
  +\frac{N}{2} -\frac{N}{2}    \frac{\overline{n}\cdot 
  (p+p^{\prime})}{\overline{n} \cdot q} \ln \frac{\overline{n} \cdot 
  p^{\prime}}{\overline{n} \cdot p} \Bigr)
n\cdot A_n  \frac{\overline{\FMslash{n}}}{2} \xi_n.  
\end{equation}

This is one definite example to show that the radiative corrections of
$O_4$ mix with $O_1+O_3$ at leading order in $\lambda$ and there appear
additional nonlocal operators. However if we consider
the radiative corrections of $O_1+O_2$, the logarithms that appear in
Eq.~(\ref{eq:nmu}) and the logarithms that appear in renormalizing the
operator $O_3$ cancel each
other and the result is given by Eq.~(\ref{sum}). We can conclude that
the reparameterization invariance 
and the residual energy invariance allow only the combination
$O_1+O_2$ at leading order in $\lambda$, and it is not renormalized at
one loop. This fact has been used in calculating the form factors for
heavy-to-light currents to order $\lambda$ in Ref.~\cite{chay1}.

We have shown explicitly that the single-gluon vertex obtained from
$O_1+O_2$ at leading order
in $\lambda$ is not renormalized at one loop. We can extend this
argument to the whole Lagrangian to all orders in $\alpha_s$. If we
require gauge invariance, reparameterization invariance and residual
energy invariance, the only possibility which includes the
interaction we have considered, is $O_1+O_2$. When we regularize both
the ultraviolet and the infrared divergences with dimensional
regularization, any loop diagrams in both theories vanish since
there is no scale involved, and the
divergence structure is the same in the full theory and in the
effective theory to all orders. Therefore the whole Lagrangian is not
renormalized to all orders in $\alpha_s$. However there is a caveat to
claim this. The collinear effective theory is a nonlocal field theory
in the coordinate $n\cdot x$ or momentum $\overline{n} \cdot p$, but
is local in other coordinates. It is possible to induce nonlocal
operators which are not present in the tree-level Lagrangian through
radiative corrections. With
this in mind, we can say that the effective Lagrangian is
not renormalized to all orders provided that no other nonlocal
operators are induced by radiative corrections. We expect that the
power counting method in Ref.~\cite{bauer7} will offer a clue to see
if nonlocal operators can exist, which are not present in the original
Lagrangian.

Note that the situation is quite different in the HQET. For the
renormalization of the chromomagnetic operator, the infrared
divergences cancel in the matching, but the ultraviolet
behavior in HQET and in the full theory is different because there is
a heavy quark mass in the full theory, while 
there is no scale in the HQET. Furthermore, if we use the on-shell
wave function renormalization, 
the wave function renormalization in the full QCD and in the HQET is
different. Therefore there appears nontrivial matching condition for
the chromomagnetic operator in the HQET. However, that is not the
case in the collinear effective theory. The difference lies in the
fact that we are dealing with massless quarks and gluons both in the
full QCD and in the effective theory. 

We have shown that the effective Lagrangian is not renormalized to all
orders in $\alpha_s$ and to all orders in $\lambda$ due to the
reparameterization invariance and the residual energy invariance
combined with the collinear and the usoft gauge invariance. This is a
very strong constraint imposed 
on the collinear effective theory, and it simplifies higher-order
corrections in $\lambda$. The Wilson coefficients of various operators
at higher order in $\lambda$ are the same as those operators at
leading order in $\lambda$. Therefore we can save a lot of
calculations in evaluating higher-order corrections to matrix elements
of some operators, and the renormalization behavior of the Wilson
coefficients of higher-dimensional operators is related to that of the
Wilson coefficients of the operators at leading order. 

We have considered the collinear sector which involves collinear
quarks, collinear gluons, and usoft gluons in the effective
theory. What we have not considered here is the soft sector and the
usoft sector, which have different symmetry 
structure. For example, there is no such
symmetry as reparameterization invariance in the usoft sector. The
usoft sector can be described rather well by the symmetries of the
full QCD. Symmetry structure and power corrections for the currents
composed of collinear fields, soft fields and usoft fields will be
useful in exclusive and inclusive decays of heavy mesons and other
high energy processes.

\section*{Acknowledgments}
The authors are supported in part by Hacksim BK21 Project and by Korea
University.

\end{document}